\documentclass[4paper,11pt]{article}
\pdfoutput=1 % if your are submitting a pdflatex (i.e. if you have
\usepackage{siunitx}
\usepackage{jinstpub} % for details on the use of the package, please
\title{\boldmath The MuPix Telescope: A Thin, high Rate Tracking Telescope}

\author[a]{H.~Augustin}
\author[b]{N.~Berger}
\author[a]{S.~Dittmeier}
\author[a]{C.~Grzesik}
\author[a]{J.~Hammerich}
\author[b]{Q.~Huang}
\author[a,1]{L.~Huth,\note{Corresponding author.}}
\author[a]{M.~Kiehn}
\author[b]{A.~Kozlinskiy}
\author[a]{F.~Meier}
\author[c]{I.~Peri\'c}
\author[a]{A.-K.~Perrevoort}
\author[a]{A.~Sch\"oning}
\author[b]{D.~vom Bruch}
\author[b]{F.~Wauters}
\author[a]{D.~Wiedner}

\affiliation[a]{Physikalisches Institut der Universit\"{a}t Heidelberg, INF 226, 69120 Heidelberg, Germany}
\affiliation[b]{Institut f\"{u}r Kernphysik, Johann-Joachim-Becherweg 45, Johannes Gutenberg-Universit\"{a}t Mainz, \newline55128 Mainz, Germany}
\affiliation[c]{Institut f\"{u}r Prozessdatenverarbeitung und Elektronik, KIT,\\ Hermann-von-Helmholtz-Platz 1, 76344 Eggenstein-Leopoldshafen, Germany}

\emailAdd{huth@physi.uni-heidelberg.de}

\abstract{The MuPix Telescope is a particle tracking telescope, optimized for
  tracking low momentum particles and high rates. 
It is based on the novel {\bf{H}}igh-{\bf{V}}oltage {\bf{M}}onolithic {\bf{A}}ctive {\bf{P}}ixel {\bf{S}}ensors (HV-MAPS), designed for the Mu3e tracking detector.
The telescope represents a first application of the HV-MAPS technology and
also serves as test bed of the Mu3e readout chain. The telescope consists of up
to eight layers of the newest prototypes, the MuPix7 sensors, which send 
data self-triggered via fast serial links to FPGAs, where the data is time-ordered and
sent to the PC. A particle hit rate of \SI{1}{MHz} per layer could be processed.
Online tracking is performed with a subset of the incoming data.
\newline\noindent
The general concept of the telescope, chip architecture, readout concept and online reconstruction are described.
The performance of the sensor and of the telescope during test beam measurements are presented.}


\arxivnumber{1234.56789} % only if you have one

\proceeding{TWEPP-16: Topical Workshop on electronics for particle Physics\\
  26.- 30. Sep.2016\\
  Karlsruhe, Germany}

\begin{document}

%\setpagewiselinenumbers
%\linenumbers
\maketitle

\flushbottom
%%% INTRO
\section{Introduction}
\label{sec:intro}

The Mu3e Experiment \cite{RP} will search for the lepton flavour violating decay of a positive muon into two positrons and one electron with a target sensitivity of 1 in 10$^{16}$~decays. 
A high rate beam  of 10$^{9}$~muons/s will be stopped on a passive target. The momentum of the decay particles as well as the vertex will be measured with a thin four layer pixelated detector. 
The high rate of low momentum particles (p~<~53\,MeV/c) requires a new detector
technology: High-Voltage Monolithic Active Pixel Sensors (HV-MAPS), combing the advantages of thin MAPS with the fast charge collection of classical hybrid pixel sensors are chosen for Mu3e.
\newline
To integrate the HV-MAPS into a multilayer tracking device, test the
 Mu3e readout architecture and for test beam 
characterization of the MuPix sensor prototypes, a tracking telescope has been
build which is discussed in the following.

\section{Telescope Concept}
\subsection*{General Concept}
\begin{figure}[htbp]
\centering
\includegraphics[width=\textwidth]{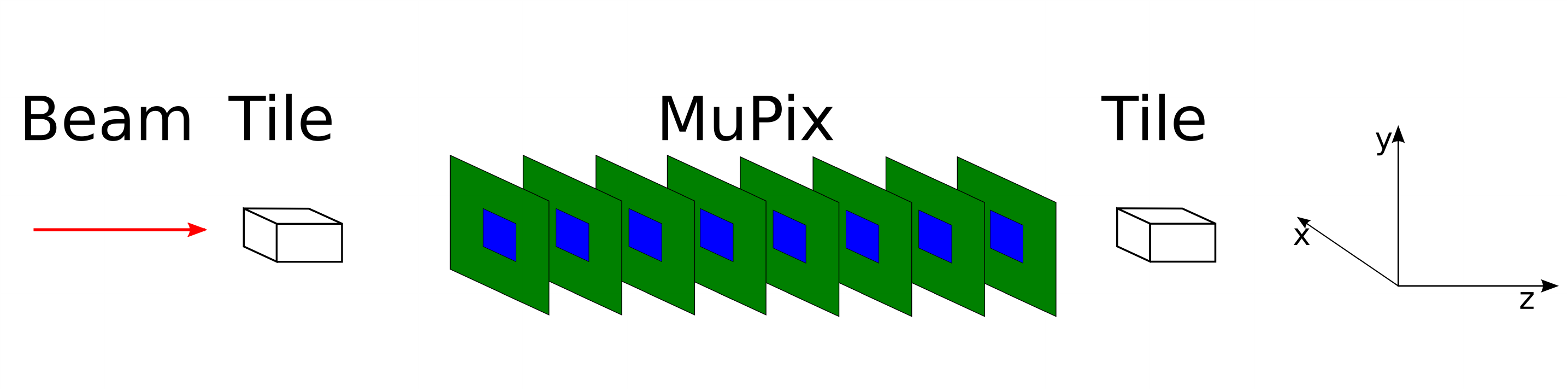}
\caption{\label{fig:concept}  Sketch of the telescope.}
\end{figure}

The MuPix telescope~\cite{Huth2014} consists of up to eight stacked layers of MuPix7 prototypes
framed by two scintillating tiles for precise reference timing, see figure~\ref{fig:concept}.
A group of four sensors is controlled by an FPGA.
The FPGAs receive the data stream from the sensors and transfer it via
PCIe to the local memory of the PC.
On the PC, data quality checks and fast online track reconstruction is performed, see section~\ref{sec:DAQ}.  The data is stored on hard
disks for offline analysis.

\subsection*{Clock and Reset Distribution}
\begin{figure}[htbp]
\centering
\includegraphics[width=.8\textwidth]{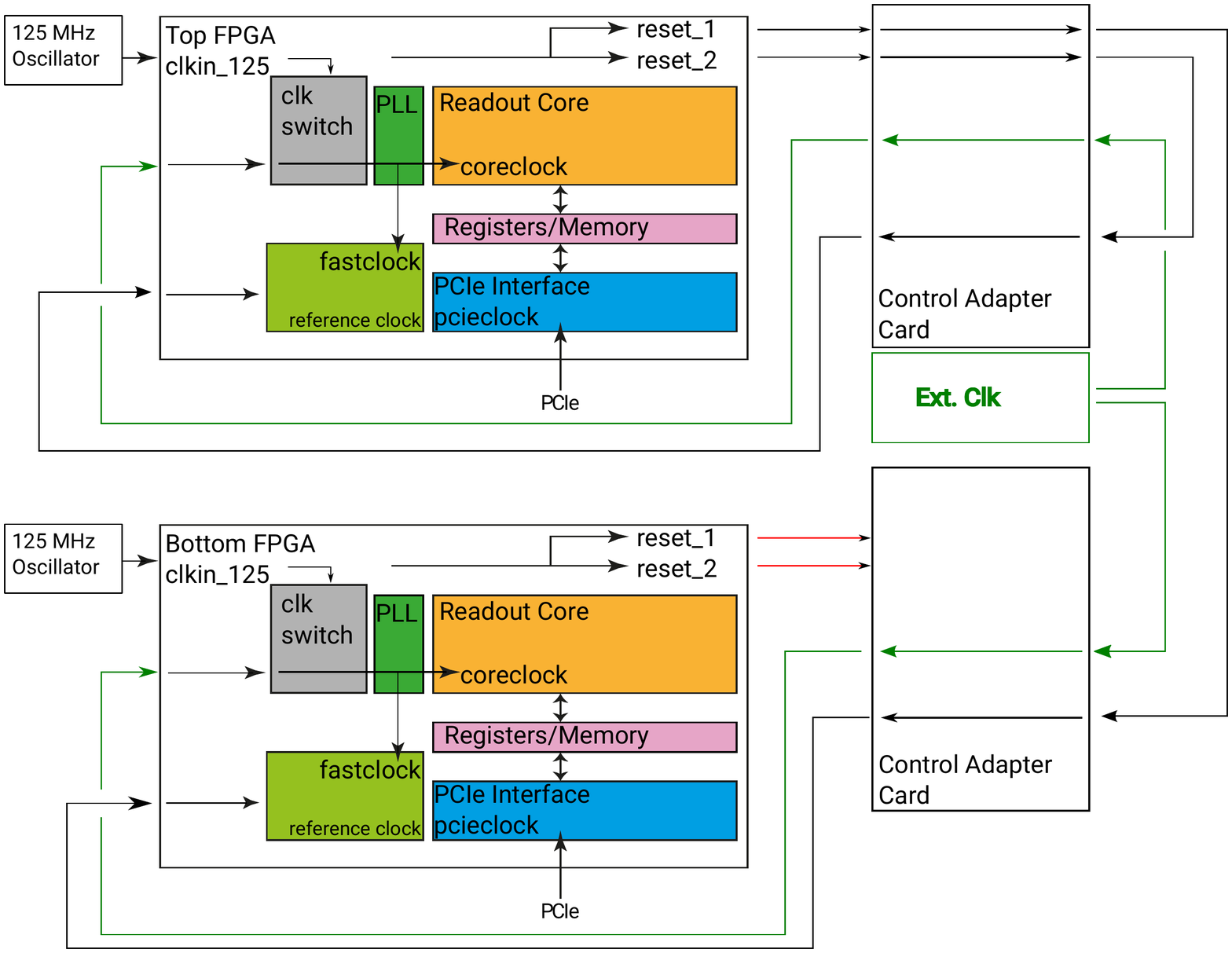}
\caption{\label{fig:clk} Block diagram of the clock and reset distribution for a setup with two FPGAs. The upper one serves as master and distributes the synchronous reset. The clock has to be provided externally. }
\end{figure}
To synchronize the data streams and time stamps from all sensors in the system,
a synchronous clock running at \SI{125}{MHz} is used. 
Due to the limited number of differential in- and outputs at the used FPGA
development board (Altera Stratix IV development kit), the core
clock for the FPGAs is applied externally, see figure~\ref{fig:clk}.
A switch on the FPGA is used to switch between the external clock 
and the internal oscillator.
A PLL on the FPGA locks to the selected clock and distributes
this clock to all connected MuPix sensors. 
Incoming signals from the tiles are sampled with a \SI{500}{MHz} counter derived from the selected clock.
To synchronize the counters a common synchronous reset signal is used.
This signal is created on the master FPGA. A loopback and a differential reset
signal
to the slave FPGA guarantee a synchronous reset of all counters.

\subsection*{Mechanical Setup}
The mechanics are based on optomechanical components from Thorlabs\texttrademark, providing a
stable and easy-to-use base system.
Two parallel rails, allowing for adjustments along the beam axis and a rotational stage for one
plane allow for flexible positioning of the sensors. 
A fully custom, aluminium Printed Circuit Board (PCB) holder mounted on movable posts on the rails allows for
fine sensor adjustments perpendicular to the beam axis with a precision of \SI{10}{\micro\meter}.
This precision is needed to correct for small sensor placing  variations between
different PCBs and therefore maximize the in beam overlap of the sensors. 
The same holder system carries the PCBs for the tiles and provides mechanical stability for
the complete system.
\newline
The Mupix chips are directly glued to a PCB, which is thinned to \SI{100 }{\micro\meter} underneath the chip to reduce scattering effects at test beams. 
The resulting total radiation length is approximately \SI{0.2}{\percent} X$_0$ per layer, assuming \SI{50}{\micro\meter} of glue.
\section{High Voltage Monolithic Active Pixel Sensors}
\begin{figure}[htbp]
\centering
\includegraphics[width=.6\textwidth]{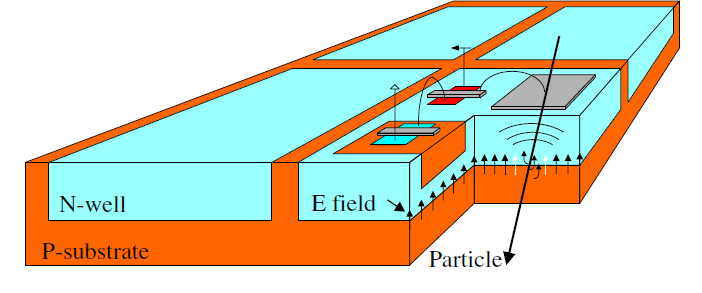}
\caption{\label{fig:hvmaps} Schematic view of the working principle of a HV-MAPS
  \cite{Peric:2007zz}. }
\end{figure}
\noindent High Voltage Monolithic Active Pixel Sensors \cite{Peric:2007zz} are
produced in a commercial \SI{180}{nm} HV-CMOS process. 
They allow for direct circuit implants into the pixel matrix.
Applying a high voltage of up to \SI{-85}{\volt} between the
n-wells and the p-substrate creates a thin depletion zone of about \SI{15}{\micro\meter}, see figure~\ref{fig:hvmaps}.
In the depletion zone, ionizing particles create electron-hole pairs, which
drift to the electrodes and create a fast charge signal, which is amplified in the pixel cell. 
The thin active depletion zone allows for removal of the inactive bulk
material. Thinning of the sensors down to
\SI{50}{\micro\meter} leads to a material budget of  \SI{0.05}{\percent} X$_0$.
The signal processing, i.e. discrimination and digitization is
implemented in a small inactive part at the edge of the sensor.
No further readout ASIC is needed which reduces complexity and costs of
the pixel detector.
\subsection*{MuPix7 Prototype}
The pixel sensor used in the MuPix telescope is the MuPix7 prototype \cite{Wiedner:2016:vienna}. 
It consists of a \SI[parse-numbers=false]{32\times40}{} pixel matrix with a pixel size of \SI[parse-numbers =
  false]{103\times80}{\micro\meter\tothe{2}}. The total active area is \SI{10.55}{mm\squared}. 
Each pixel cell features an integrated amplifier and has a point to point
connection to its partner cell in the periphery of the sensor.
In each digital cell, the signal is compared to a threshold, which can be
fine-adjusted with tune digital to analog converters (TDACs) for each pixel. The cell address and an 8bit time stamp, running at \SI{62.5}{MHz} are stored.
An integrated finite state machine performs a priority readout of the hits
and sends them 8bit/10bit encoded over a 1.25 GBit/s serial link to the readout FPGA.  
%
%*********** DAQ SYSTEM**************
\section{DAQ System}
\label{sec:DAQ}
\begin{figure}[htbp]
\centering
\includegraphics[width=.9\textwidth]{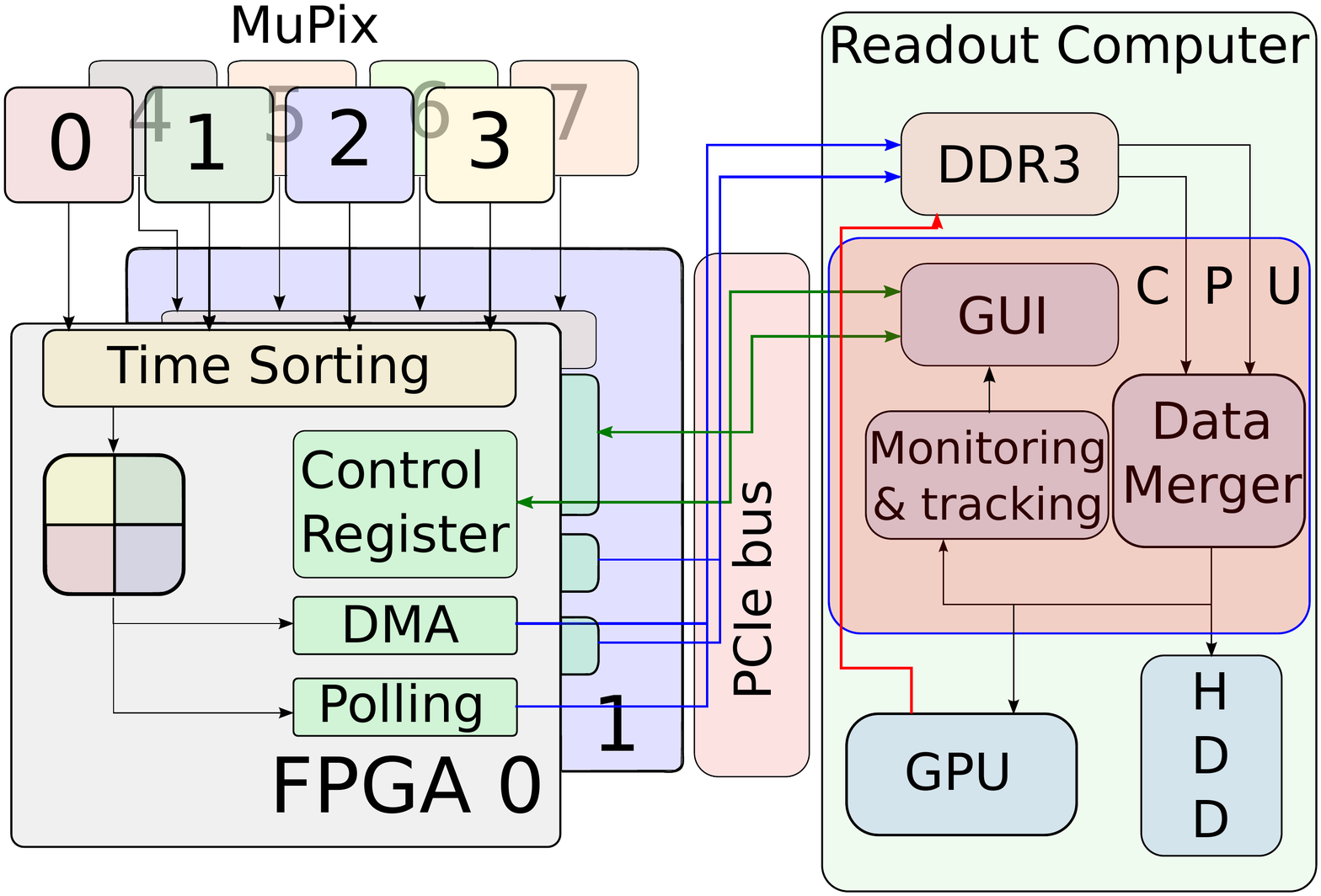}
\caption{\label{fig:daq} Block diagram of the data and control flow in
  the DAQ system for the MuPix telescope. The trigger inputs are not displayed.}
\end{figure}
\noindent The DAQ system is designed to run on a single computer with powerful GPU and
CPU (i7-3930K @~\SI{3.2}{GHz}, GeForce~GTX~680).
The sensor control and readout are steered over two  FPGA boards, which are
connected to the computer via the PCIe bus. 
Each FPGA controls and receives data from four MuPix7 sensors and is able to receive signals from the scintillating tiles.
The FPGAs can be controlled via a graphical user interface (GUI) on the computer.
The GUI also provides access to the data quality checker and online tracking
results.
A block diagram of the data flow is shown in figure~\ref{fig:daq}.
\newline
The incoming serial hit data from the four sensors is decoded online.
Consequently, a sensor label is assigned to each hit and the four data streams
are merged, sorted by their time stamps and assigned to blocks with 32 time stamps each.
In parallel, blocks with tile signals are created. 
The blocks are copied to the local memory of the computer.
\newline
The data transfer from the FPGA to the PC was first realized using polling, where the
CPU requests data from the FPGA.
This introduces large overhead and delay in the copying
process.
Therefore, fast direct memory access (DMA) is also implemented for operation at very high rates.
Using DMA, the system bottleneck is the disk writing speed, while in the
polling mode, the polling request overhead limits the readout speed.
Using polling, a track rate of 1 MHz is achieved without errors. 
For the DMA, no rate limit is determined up to now. 
The theoretical limit is high enough to transfer the data from eight MuPix7
sensors at full rate of approximatly \SI{30}{MHits/s} per sensor.
%But there is no possibility to store this amount of data on disk.
\newline 
The received data from the two FPGAs is merged into a common data stream and
written to disk. 
The data, which is kept in the RAM of the computer is queued to a data quality
checker, which executes online monitoring tasks.
The time stamp distributions, hit maps, time and spatial correlation plots provide
control over the system performance, ensuring timely feedback and high efficiency in test beam
measurements.
The sensor efficiencies and noise rates are also directly computed and
displayed.
Data frames with a potential track, i. e. hits in all layers are further queued
to an online tracking thread, which returns tracking residuals. They are also displayed in the GUI.
This allows for quick alignment and continuous performance control.
\newline
To test the performance of the GPU online tracking, the data can be
copied to the GPU, which performs the same tracking operations as the CPU. 
Only the residuals for the calculated tracks are returned. It was found, that
the GPU tracking provides exactly
the results as expected \cite{Grzesik2016}.
In addition, online tracking performance tests with DMA from the FPGA to the
RAM and then to the GPU have been successfully
performed.
%The DMA from the FPGA over the local RAM to the GPU into the full readout chain is not implemented so far.
\newline
For the online track reconstruction, a straight line fit is implemented. It can be
analytically solved and does not require any iterative procedure.
This makes the fit very fast and robust. 
It also has the advantage that the number of calculations is known
and the workload on the GPU can be optimized.
%********TESTBEAM RESULTS**********
%
\section{Testbeam Results}

Several telescopes have been used at DESY (\SI{4}-\SI{6}{GeV/c} e$^+$), PSI
(\SI{150}-\SI{250}{MeV/c} p$^+$, $\pi^+$, e$^+$), MAMI (\SI{950}{MeV/c} e$^+$) and
CERN (\SI{180}{GeV/c}  $\pi^+$) test beam campaigns. The rates varied from several
kHz to \SI{1}{MHz}. The telescope performed well at the different beam conditions.
The DAQ system handled track rates of up to 1 MHz and
was used as a crucial tool for MuPix characterization measurements.
\subsection*{Alignment and Stability}
\begin{figure}[htbp]
\centering
\includegraphics[width=.47\textwidth]{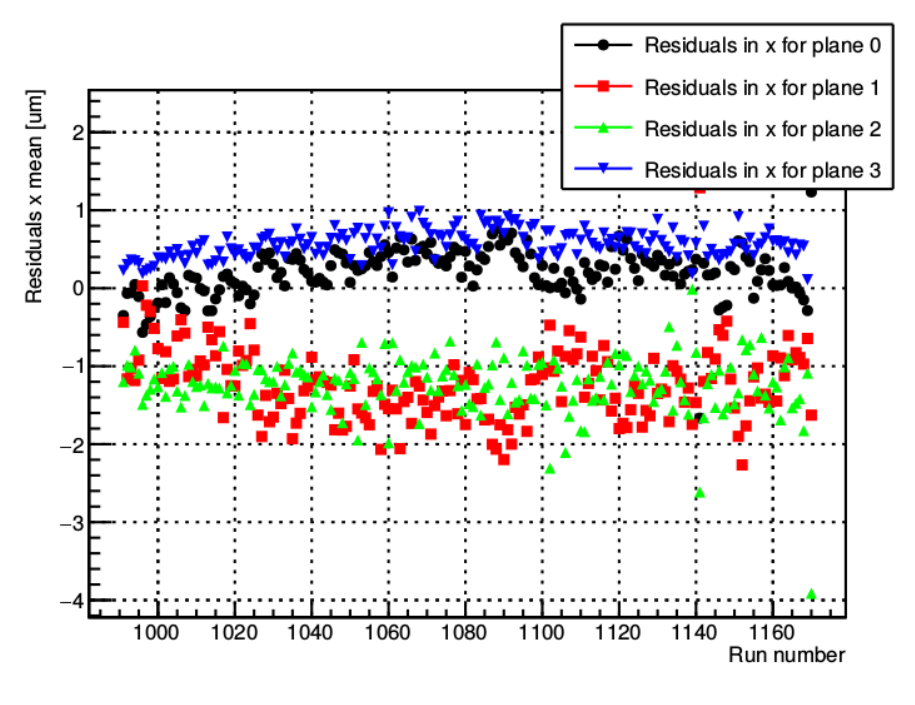}
\qquad
\includegraphics[width=.47\textwidth]{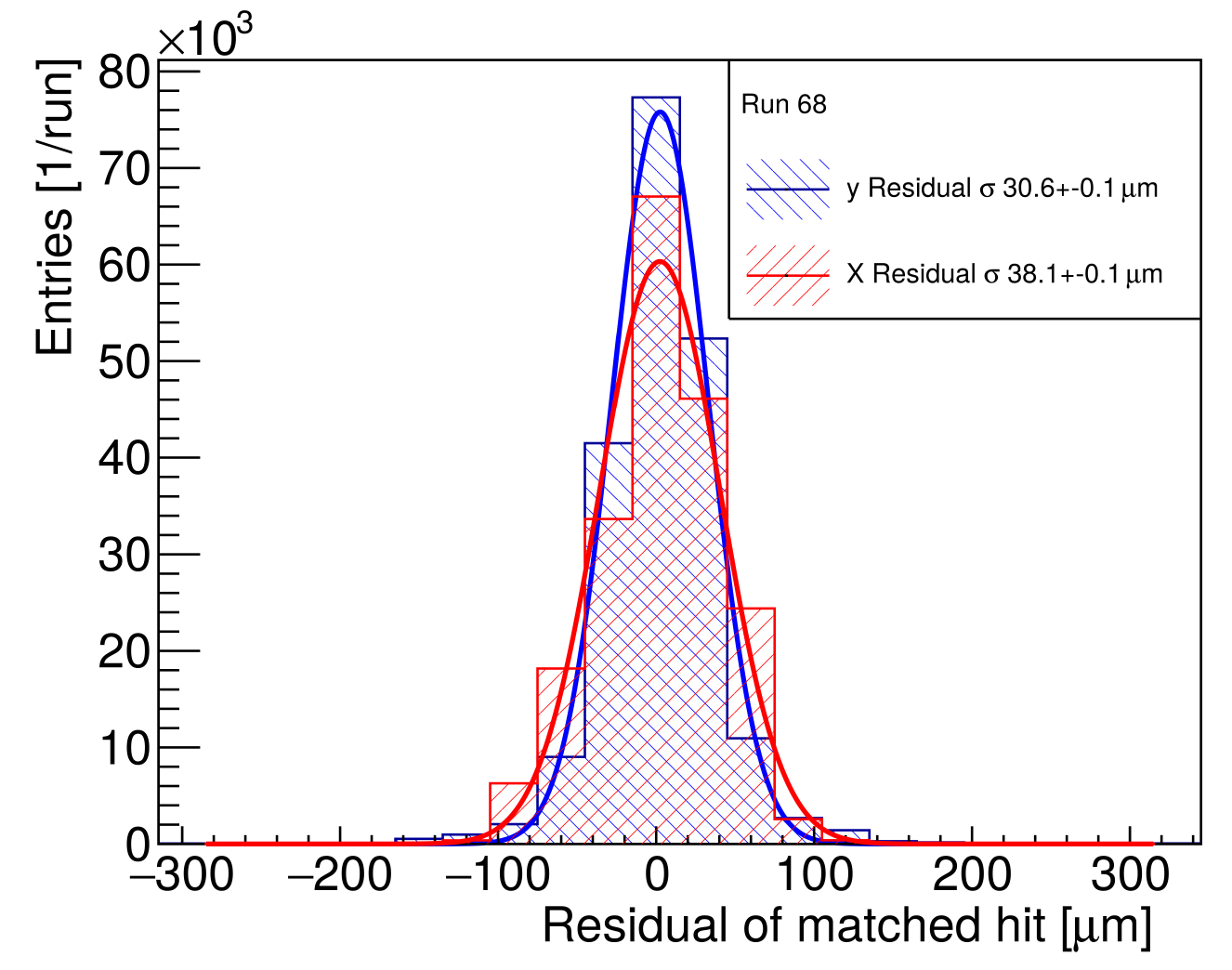}
\caption{\label{fig:alignment}Left: Mean residuals of one set of runs at DESY. The
  time period covered is about 10 hours. Right: Residuals of a MuPix7 sensor at DESY
  with 4 GeV positrons.}
\end{figure}
The mechanical alignment precision is of the order of \SI{100 }{\micro\meter}
and corrected offline via automated track based alignment to less than \SI{5}{\micro\meter}, see figure~\ref{fig:alignment}. 
The variation of the mean value of the residuals stays below \SI{2}{\micro\meter} during the displayed time period of 10~hours.
%There is no significant change in the residuals during the covered 10~hours
This proves the mechanical stability.
The MuPix7 spatial resolution, measured with the MuPix telescope is
shown in figure~\ref{fig:alignment} and follows the expected behavior. The
spatial resolution is given by the combination
of the limited pixel size and the scattering on the planes, which is not included in
the fit.
For the analyzed run the layer spacing is approximately \SI{3}{cm} and the
sensors are not rotated. Possible unintended rotations are not taken into account for the alignment.
The slight difference of the residual width between x and y resolution is caused by the different
dimensions of the pixels (\SI[parse-numbers = false]{103\times80}{\micro\square\meter}).
\subsection*{Efficiency, Noise and Time Resolution}
\begin{figure}[htbp]
\centering
\includegraphics[width=.47\textwidth]{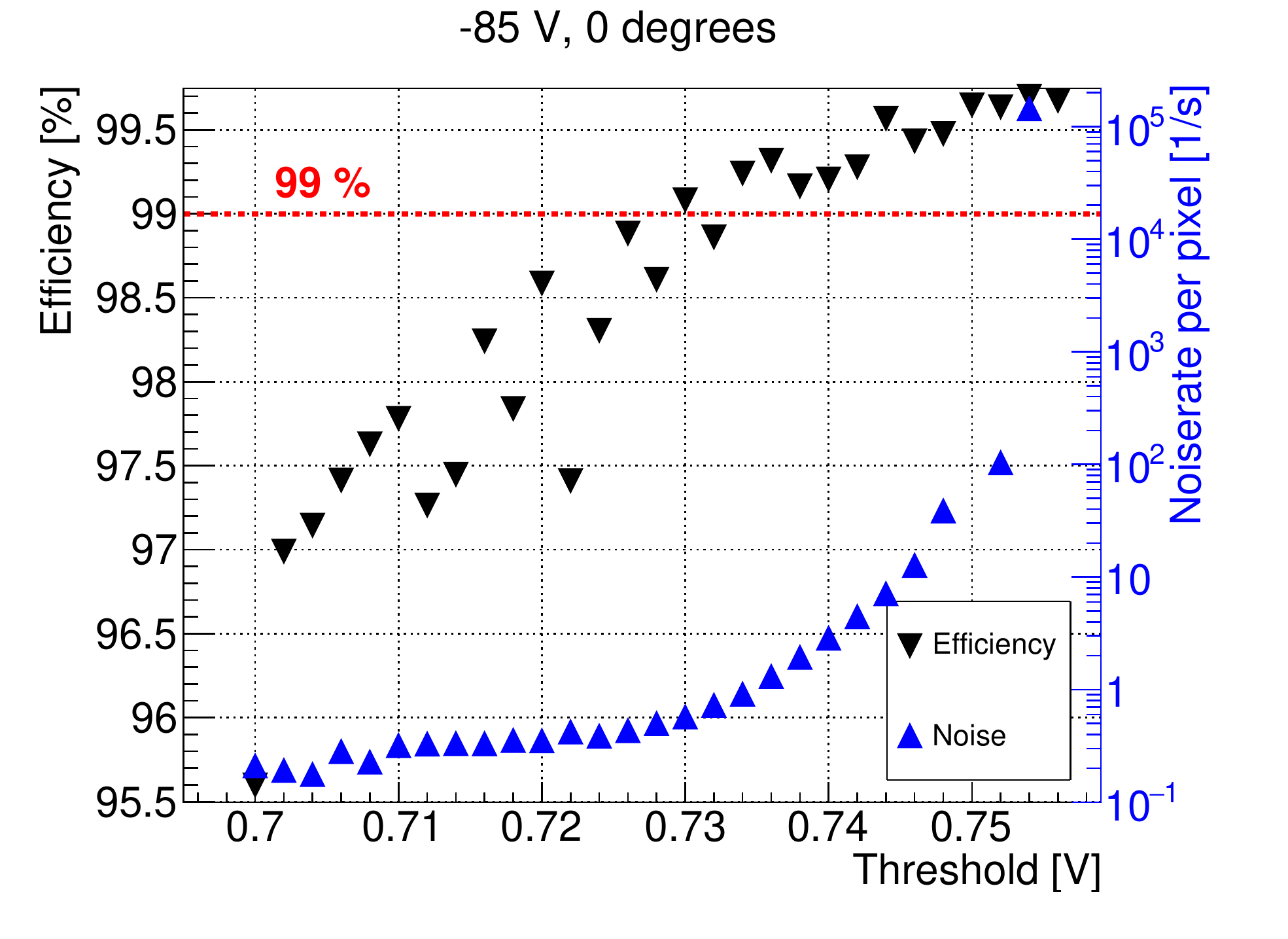}
\qquad
\includegraphics[width=.47\textwidth]{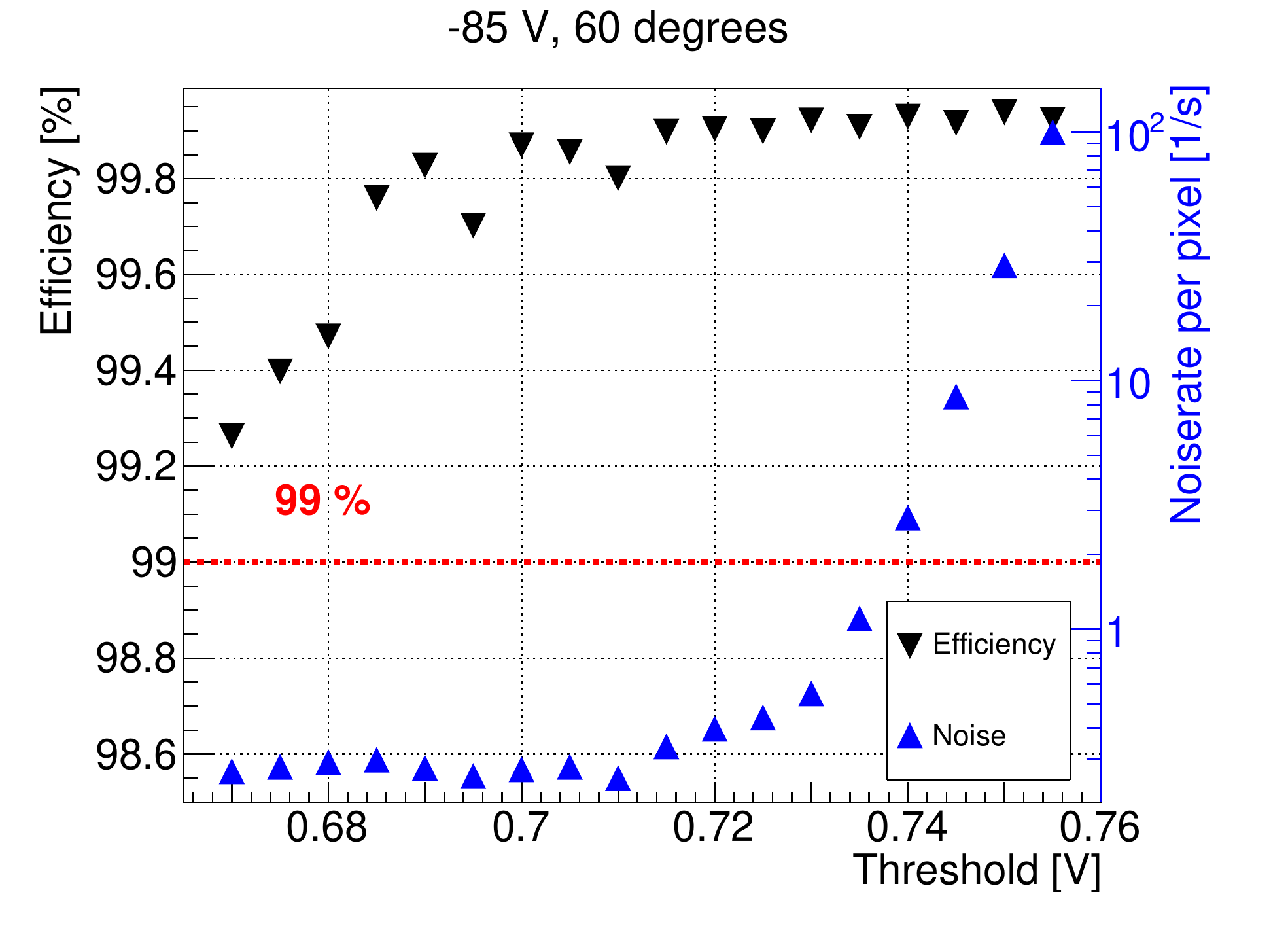}

\caption{\label{fig:efficiency}Efficiency and noise for perpendicular tracks
    %with a search window of \SI{800}{\micro\meter} and a time window of $\pm$
    %\SI{48}{\nano\second}. 
  The statistical errors are smaller than the markers. Left: Perpendicular sensor.  Right: sensor rotated by
    \SI{60}{\degree}.}
\end{figure}

\begin{figure}[htbp]
\includegraphics[width=\textwidth]{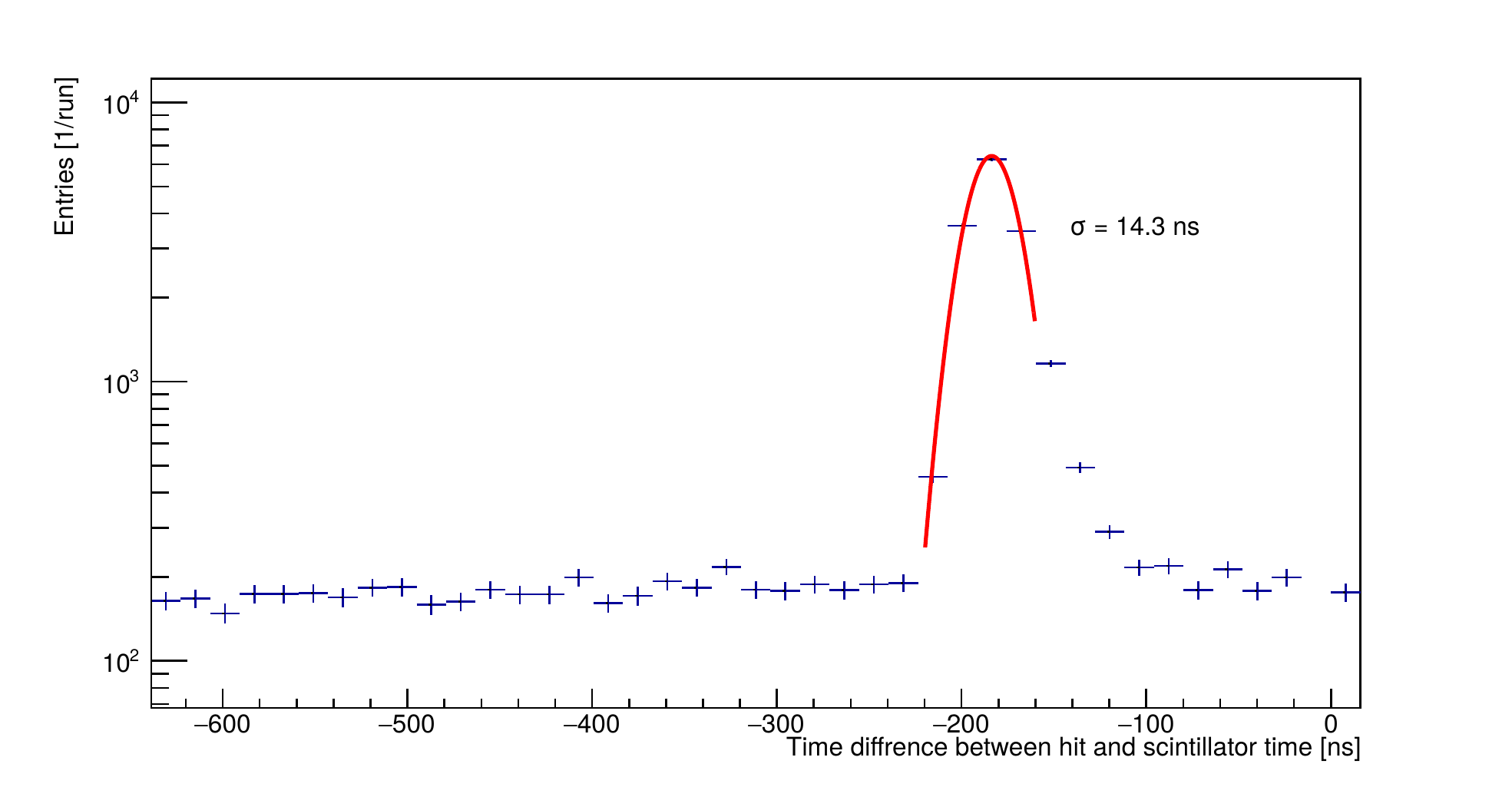}
\caption{\label{fig:timing} Time resolution of the MuPix7, relative to the scintillating tiles.	}
\end{figure}

%To operate the Mu3e pixel detector, hard constraints are set on the efficiency, noise and time resolution of the pixel sensor. 
%We require a time resolution better than \SI{20}{ns}, efficiency above \SI{99}{\percent} with less than \SI{20}{\hertz}
%noise per pixel. 
%Due to cooling limitations with gaseous helium for the Mu3e detector, the power
%consumption has to stay below \SI{400}{ mW/cm\tothe{2}}.
%\newline 
MuPix7 prototypes were tested at the DESY test beam using settings corresponding to a power
consumption per area of \SI{300}{mW/cm\tothe{2}}.
%fulfilled all the requirements.
For track matching a circular search window of \SI{800}{\micro\meter} around the
extrapolated track intersection and a time window of $\pm$~\SI{48}{ns} around
the track time is chosen.  
An efficiency above \SI{99}{\percent} is reached for perpendicular tracks
at a threshold of \SI{730}{\mV}, while the noise stays below the \SI{20}{\hertz} noise per pixel limit up to
\SI{745}{mV}, see figure~\ref{fig:efficiency}.
Therefore, a threshold operation range of \SI{15}{mV} can be used.
This can be improved by rotating the sensor, which increases the particles'
path in the depletion zone.
The higher charge  deposit leads to a larger signal and better performance. 
A sensor rotated by  \SI{60}{\degree} shows a threshold operation
range from \SI{680}-\SI{745}{mV}, see figure~\ref{fig:efficiency}.
A similar increase in signal size as for the tilt can be achieved by using a substrate with higher
resistivity, which is planned for the upcoming chip submission.  
\newline
The time resolution of the MuPix7 is studied using the tile scintillators as
precise reference. Events with one track and one tile signal are selected. The
resulting time resolution averaged over all pixels  is \SI{14.3}{ns}, expressed as Gaussian~$\sigma$, see figure~\ref{fig:timing}.
%********** CONCLUSION ********
\section{Conclusion and Outlook}
The MuPix telescope is used for MuPix sensor characterization and
provides a perfect framework for the development of the Mu3e pixel detector.
\newline
The telescope was successfully operated at different beam environments. The
highest  useful particle rate is of the order of 1 MHz running the full readout chain.
\newline
The MuPix telescope is  used to show that the MuPix7 fulfils all performance
goals, required for a successfull operation of the Mu3e detector: Efficiencies
above \si{99 \percent} at moderate noise rates below \SI{20}{Hz} per pixel and time resolution of \SI{14.3} {ns}. 
The estimated power consumption for these measurements is \SI{300}{mW/cm\tothe{2}}.
\newline
The MuPix telescope will be equipped with the new MuPix8 prototype and further used
to characterize the sensors. 
%In addition, it is also planned to provide
%reference tracks for other systems, like the Mu3e fibre detector.

\section*{Acknowledgements}
N.~Berger, Q.~Huang, A.~Kozlinskiy, D.~vom~Bruch and F.~Wauters
thank the \textit{Deutsche Forschungsgemeinschaft} for supporting them and the
Mu3e project through an Emmy Noether grant. 
S.~Dittmeier and L.~Huth acknowledge support by the IMPRS-PTFS. 
A.-K.~Perrevoort acknowledges support by the Particle Physics beyond the
Standard Model research training group [GRK 1940].
H.~Augustin acknowledges support by the HighRR research training group [GRK
  2058]. 
N.~Berger and A.~Kozlinskiy thank  the PRISMA Cluster of Excellence for
support. 
\newline
The measurements leading to these results have been performed at the Test Beam
Facility at DESY Hamburg (Germany), a member of the Helmholtz Association
(HGF).
We thank the Institut f\"{u}r Kernphysik at the Johannes Gutenberg University
Mainz for giving us the opportunity to take data at the MAMI beam. 
We would like to thank PSI for valuable test beam time.
We owe our SPS test beam time to the SPS team and our LHCb colleagues,
especially Heinrich Schindler, Kazu Akiba and Martin van Beuzekom.

\bibliography{mybibfile.bib}

\end{document}